\documentclass{elsart}
\usepackage{amsmath,amssymb}
\usepackage{mathrsfs}
\usepackage{graphicx}
\usepackage[square,comma,numbers,sort&compress]{natbib}

\newcommand{\ave}[1]{\left \langle {#1} \right \rangle}
\newcommand{\abs}[1]{\left \lvert {#1} \right \rvert}
\newcommand{\ket}[1]{\left \lvert {#1} \right \rangle}
\newcommand{\kB}{k_{\mathrm{B}}}
\newcommand{\fano}{\mathscr{F}}
\newcommand{\spectral}{\mathscr{S}}

\journal{Optics Communications}

\begin{document}

\begin{frontmatter}

\title{Two-level laser light statistics}

\author[cem2]{L. Chusseau\corauthref{cor}\thanksref{support}}%
\corauth[cor]{Corresponding author.}%
\ead{chusseau@univ-montp2.fr}%
\ead[url]{www.opto.univ-montp2.fr/\string~chusseau}%
\address[cem2]{Centre d'{\'E}lectronique et de
Micro-opto{\'e}lectronique de Montpellier, UMR 5507 CNRS,
Universit{\'e} Montpellier II, 34095 Montpellier, France}%
\thanks[support]{This work was supported by the STISS Department of
Universit{\'e} Montpellier II and by CNRS under the JemSTIC Program.}

\author[liron]{J. Arnaud}%
\ead{arnaudj2@wanadoo.fr}%
\address[liron]{Mas Liron, 30440 Saint Martial, France}

\author[lirmm]{F. Philippe\thanksref{also}}%
\ead{fabrice.philippe@univ-montp3.fr}%
\address[lirmm]{Laboratoire d'Informatique de Robotique et de
Micro{\'e}lectronique de Montpellier, UMR 5506 CNRS, 161 Rue Ada,
34392
Montpellier, France}%
\thanks[also]{Also at MIAp, Universit{\'e} Paul Val{\'e}ry, 34199
Montpellier, France}

\date{\today}

\begin{abstract}
The statistics of the light emitted by two-level lasers is evaluated on the basis of generalized rate equations.  According to that approach, all fluctuations are interpreted as being caused by the jumps that occur in active and detecting atoms. The intra-cavity Fano factor and the photo-current
spectral density are obtained analytically for Poissonian and quiet pumps.  The algebra is simple and the formulas hold for small as well as large pumping rates. Lasers exhibit excess noise at low pumping levels.  
\end{abstract}
		
\end{frontmatter}

\section{Introduction}

We consider the light statistics of two-level atom lasers.  Analytical expressions for the photo-detection spectral density, $\spectral (\Omega )$, and the Fano factor, $\fano \equiv \mathrm{var}(m)/\ave{m}$, where $m$ denotes the number of photons in the laser cavity, $\mathrm{var}(m)$ the variance of $m$, and $\ave{m}$ the average value of $m$, are derived.  Numerous paper have addressed this problem before, e.g. \cite{yamamoto:PRA86, kennedy:PRA89, khazanov:PRA90, ritsch:PRA91, ralph:PRA91, kolobov:PRA93, levien:PRA93,
koganov:PRA00}.  However, explicit formulas that are not restricted to high-pumping levels and do not neglect spontaneous decay do not seem to be available.  The concepts on which rests our generalized rate-equations approach have been explained in detail in tutorial papers \cite{arnaud:OQE95, arnaud:OQE01} and applied to three- and four-level lasers \cite{chusseau:3levels}. This method enables one to obtain analytical expressions straightforwardly, which coincide with Quantum-Optics results whenever a comparison can be made, even in the case of sub-Poissonian light statistics. The novelty of the paper resides in the fact that a simple formula is found that holds irrespectively of the pumping rates.  Similar formulas were previously obtained for four-level lasers \cite{chusseau:3levels}, but they were too involved to be written down.  The present paper may be viewed as an illustration of that previous work, with formulas that are easier to appreciate because of their simplicity.

Rate equations treat the number of photons in the cavity as a classical random function of time.  The light field is quantized as a result of matter quantization and conservation of energy, but not
directly.  Rate equations should be distinguished from semi-classical theories in which the optical field is driven by atomic dipole expectation values.  The present theory rests instead on the
consideration of transition probabilities as in \cite{loudon}.  Every absorption event, even if it occurs in the detector, reacts on the number of photons in the optical cavity. Our method allows the use of Monte-Carlo simulations \cite{chusseau:OQE01,chusseau:3levels}.

\begin{figure}
   \begin{center}
   	\includegraphics[width=0.45\linewidth]{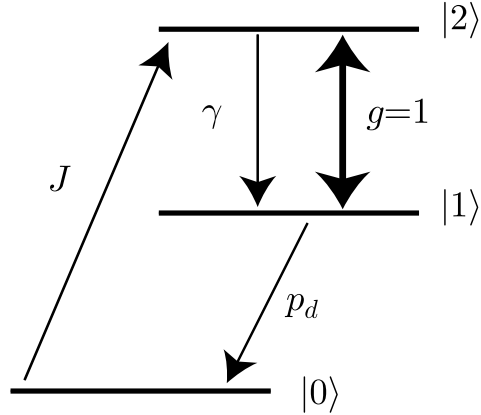}
   \end{center}
   \caption{Level scheme of a three-level laser.  The two-level laser obtains when $1/p_{d} \to 0$.}
   \label{scheme}
\end{figure}

The active medium considered is a collection of $N$ identical atoms. Transitions are supposed to be strongly homogeneously broadened so that atomic polarizations may be adiabatically eliminated.  The two-level scheme is obtained as a limit case of the three-level one (see
Fig.~\ref{scheme}), with the levels labeled $\ket{0}$, $\ket{1}$, and $\ket{2}$ in increasing order of energy.  Level separations are supposed to be large compared with $\kB T$, where $T$ denotes the atomic temperature and $\kB$ the Boltzmann constant, so that thermally-induced transitions are negligible.  The classical 2-level scheme is a valid one in the limit in which the lower decay time $1/p_{d}$ tends to zero.  In that limit, the $\ket{1}$ population tends to zero and stimulated absorption from $\ket{1}$ to $\ket{2}$ does not occur. For concreteness, we may think of the pumping as resulting from the injection of atoms in the excited state inside the cavity, as in
\citet{scully:PR67}. If the atoms are injected regularly the fluctuations of the pumping rate vanish ($\xi =0$ in our notation).

The probability per unit time that an electronic transition from level $\ket{2}$ to $\ket{1}$ occurs is taken as equal to $m+1$, where $m$ denotes the number of photons in the cavity.  This amounts to selecting a time unit for which the laser gain $g \equiv 1$. Spontaneous decay from level $\ket{2}$ to level $\ket{1}$ is allowed with probability $\gamma$.  This decay may be either non-radiative or involve radiation into other electromagnetic modes, besides the one of interest.  Photons are absorbed with probability $\alpha m$, where $\alpha$ denotes a constant, the absorbing atoms residing most of the time in their ground state.  These absorbing atoms model the
transmission of light through mirrors with subsequent absorption by a detector.  Provided detection is linear and reflectionless, it is immaterial whether absorption occurs inside or outside the optical
cavity.  For simplicity, internal absorption is neglected.

\citet{golubev:JETP84} first pointed out in \citeyear{golubev:JETP84} that if the laser pumping rate does not fluctuate and detection is ideal, the detected current should not fluctuate at low Fourier (or
baseband) frequencies.  This observation was made independently by \citet{yamamoto:PRA86}.  These latter authors furthermore proposed \citeyear{yamamoto:PRA86}{\nobreakspace} a practical way of realizing quiet electrical pumps, namely a battery in series with a cold resistance, much greater than the laser diode dynamic resistance \cite{machida:PRL87}.  This scheme was implemented for the first time in \citeyear{machida:PRL87}{\nobreakspace}.  The pumping rate is
supposed in the present paper to be fully prescribed through its average value $J$ and the spectral density $\sigma_{j}\equiv \xi J$ of its fluctuations.  It would be straightforward, however, to account for a dependence of the pumping rate on the atomic populations, as is required in the case of optical pumps \cite{chusseau:3levels}. 

We first consider the steady-state regime and subsequently derive formulas for the photo-detection spectrum and the intra-cavity field Fano factor.

\section{Steady-state}
\label{SS}

Let $n_{j}$, $j = 0, 1, 2$, denote the number of atoms in state $j$.  In the limit $1/p_{d}=0$, we have $n_{1}=0$, and thus
\begin{align}
    N = n_{0}+n_{2}.
    \label{conservation}
\end{align}

Let $\mathcal{J}$ denote the pumping rate, $\mathcal{R}$ the net stimulated rate, $\mathcal{S}$ the spontaneous decay rate from the upper to the lower working levels, and $\mathcal{Q}$ the photon absorption rate.  The steady-state conditions read
\begin{align}
\label{steadystate}
	\mathcal{J} &= \mathcal{R} + \mathcal{S} , &
	\mathcal{Q} &= \mathcal{R} ,
\end{align}
where
\begin{align}
\label{rates}
	\mathcal{J} &= J , &
	\mathcal{R} &= (m+1) n_{2} , &
	\mathcal{S} &= \gamma n_{2} , &
	\mathcal{Q} &= \alpha m .
\end{align}

Equations \eqref{conservation}, \eqref{steadystate} and \eqref{rates} provide the steady-state atomic populations $n_{i}$ and photon number $m$ as
\begin{align}
	m &=\sqrt{{\left( \widehat{J} + \gamma + 1 \right)}^{2} - 4 \widehat{J} \gamma} - \gamma - 1, &
	n_{2} &= \frac{\alpha m}{m + 1} ,&
	n_{0} &= N - \frac{\alpha m}{m	+ 1} ,
	\label{steadysol}
\end{align}
where $\widehat{J} = J /  \alpha$.

When $\gamma = 0$, Eq.~\eqref{steadysol} leads to
\begin{align}
	m &= \widehat{J} , &
	n_{2} &= \frac{\alpha \widehat{J}}{\widehat{J} + 1} , &
	n_{0} &= N - \frac{\alpha \widehat{J}}{\widehat{J} + 1} .
	\label{steadyzero}
\end{align}

\section{Photo-detection spectrum}
\label{PDS}

Within the weak-noise approximation atomic and photonic populations split into steady-state values and fluctuations.  For example the instantaneous photon number $m$ may be written as $m=\ave{m} + \Delta m$, where $\ave{m}$ needs not be not distinguished from the steady-state value of $m$.  Similarly, rates get split into steady-state values and fluctuations consisting of a deterministic function of the population fluctuations and uncorrelated Langevin
``forces''.  For example $\mathcal{J}$ splits into $J =\ave{\mathcal{J}}$ and the sum, $\Delta J$, of a deterministic function of populations and a Langevin force $j(t)$ expressing the jump process randomness.  Thus
\begin{align}
	\label{fluctuations}
	\mathcal{J} &\equiv J + \Delta J , &
	\mathcal{R} &\equiv R + \Delta R , &
	\mathcal{S} &\equiv S + \Delta S , &
	\mathcal{Q} &\equiv Q + \Delta Q .
\end{align}
where a first-order variation of the expressions in Eq.~\eqref{rates} gives
\begin{subequations}
    \label{variations}
    \begin{align}
		\Delta J &=  j , \\
		\Delta R &= \left( m + 1 \right) \Delta n_{2} + n_{2} \Delta m + r , \\
		\Delta S &= \gamma \Delta n_{2} + s , \\
		\Delta Q &= \alpha \Delta m + q .
    \end{align}
\end{subequations}

Since the number $N$ of atoms is supposed to be constant we have
\begin{equation}
    \label{conservationDelta}
    \Delta n_{0} + \Delta n_{2} = 0 .
\end{equation}

At some Fourier frequency $\Omega$, \footnote{The Fourier angular frequency is called here `frequency' for short. The frequency, as it is usually defined, is obtained by dividing $\Omega$ by $2 \pi$.} the generalized rate equations read \cite{arnaud:OQE95}
\begin{subequations}
    \label{steadyf}
    \begin{align}
	i \Omega \Delta m & = \Delta R - \Delta Q , \\
	i \Omega \Delta n_{0} & = \Delta R + \Delta S - \Delta J , \\
	i \Omega \Delta n_{2} & = \Delta J - \Delta R - \Delta S .
    \end{align}
\end{subequations}

Equations \eqref{fluctuations}--\eqref{steadyf} are solved for $\Delta Q$, yielding a linear combination of the Langevin forces
\begin{equation}
    \Delta Q = \sum_{z \in \{j,q,r,s\}} \tilde{c}_{z} \, z ,
    \label{DeltaQ}
\end{equation}
where the $\tilde{c}_{z}$ are complex and frequency dependent coefficients. The normalized photo-current spectral density is thus
\begin{equation}
	\spectral(\Omega ) = \frac{1}{\alpha m} \sum_{z \in \{j,q,r,s\} }
	\abs{\tilde{c}_{z}}^2 \, \sigma_{z} ,
    \label{SDfnonnull}
\end{equation}
where $\sigma_{z}$ denotes the spectral density value of the Langevin noise source $z$, equal to average rates
\begin{align}
	\label{SDlangevin}
	\sigma_{j} &= \xi J , &
	\sigma_{q} &= \alpha m , &
	\sigma_{r} &= \left( m + 1 \right) n_{2} , &
	\sigma_{s} &= \gamma n_{2} .
\end{align}
$\xi$ accounts for a poissonian pump if $\xi = 1$, while a perfect quiet pumping is obtained at $\xi=0$.  When these expressions are introduced in Eq.~\eqref{SDfnonnull} an analytical expression of $\spectral$ is obtained
\begin{equation}
	\spectral ( \Omega ) = 1 +\frac{\alpha}{m}\frac{\mathscr{U} ( \Omega )}{\mathscr{D} ( \Omega )},
	\label{SDOmega}
\end{equation}
with
\begin{subequations}
 	\label{UandD}
	\begin{align}
		\mathscr{U}(\Omega) &= \left( n_{2} ( 3m+1) -  \alpha m \right) \Omega^2 + n_{2} ( 3m+1) \gamma \widehat{m} - \alpha m \widehat{m}^2 + J ( m+1 )^2 \xi , \\
		\mathscr{D}(\Omega) &=  \Omega^4 +  \left( { \left( \widehat{m} + \alpha - n_2 \right) } ^2 +  2 \left( \gamma n_2 - \alpha \widehat{m} \right) \right) \Omega^2 + \left( \gamma n_{2} - \alpha \widehat{m} \right)^2  .
		\label{D}
   \end{align}
\end{subequations}
where $\widehat{m}$ stands for $m+\gamma +1$. $\spectral(\Omega) - 1$ is the ratio of a polynomium of degree 1 in $\Omega^2$ in the numerator and of degree 2 in $\Omega^2$ in the denominator.  A few observations can be made from Eqs.~\eqref{SDOmega}--\eqref{UandD}
\begin{itemize}
	\item Well above threshold and with $\gamma = 0$, implying that $m \gg 1$ and $J \gg 1$, Eqs.~\eqref{SDOmega}--\eqref{UandD} simplify to \( \spectral = { \left( \alpha^2 \xi + \Omega^2 \right) } / { \left(	\alpha^2 + \Omega^2 \right) } \), as is well-known, see e.g. \cite{arnaud:OQE95};
	
	\item $\lim_{\Omega \to \infty}\spectral(\Omega) = 1$.  That is, the light statistic is Poissonian at high Fourier frequencies irrespective of the pumping and laser parameters;

	\item $\spectral(0) = \xi$ for an ideal laser ($\gamma = 0$), i.e., laser light spectrum at zero-frequency corresponds to the value of the pump $\xi$-parameter;

	\item For usual laser parameters there is a value of $\Omega$ that maximizes $\spectral(\Omega)$, according to Eq.~\eqref{SDOmega}.
\end{itemize}

\begin{figure}
		\begin{tabular}{cc}
		(a) & (b) \\
		\includegraphics[width=0.45\linewidth]{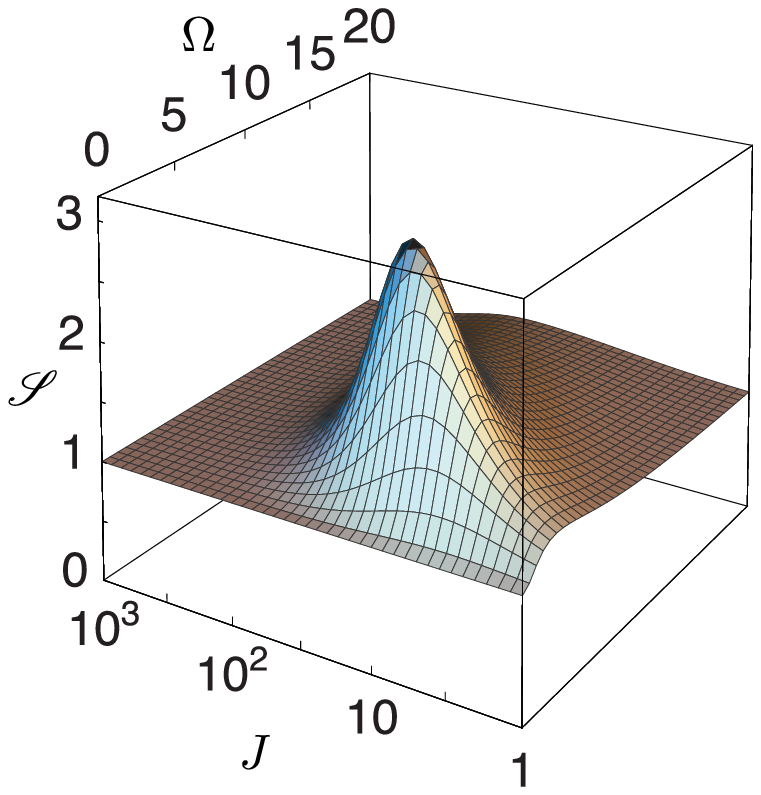} & 
		\includegraphics[width=0.45\linewidth]{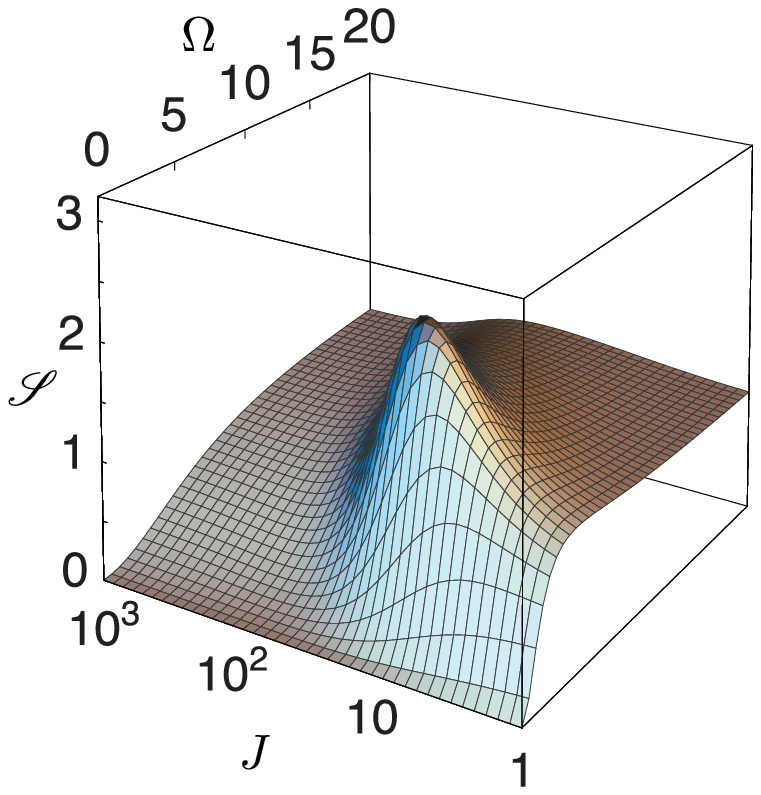} \\
		\includegraphics[width=0.45\linewidth]{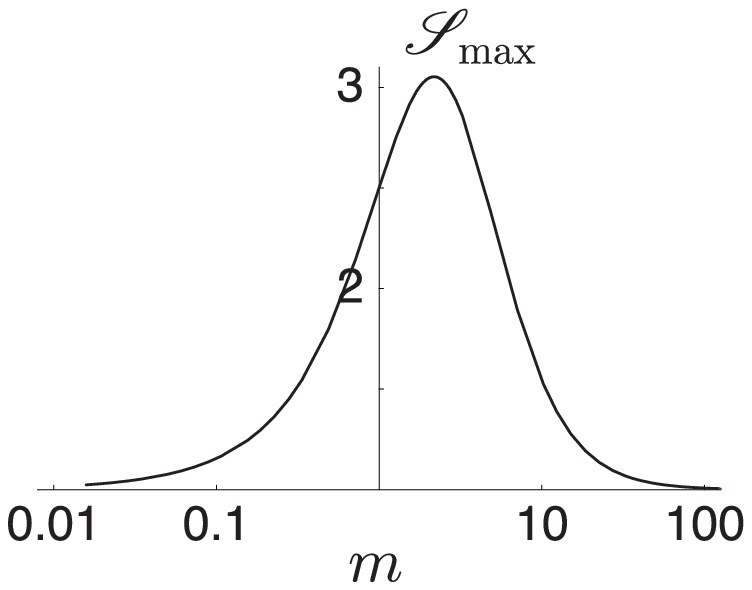} & 
		\includegraphics[width=0.45\linewidth]{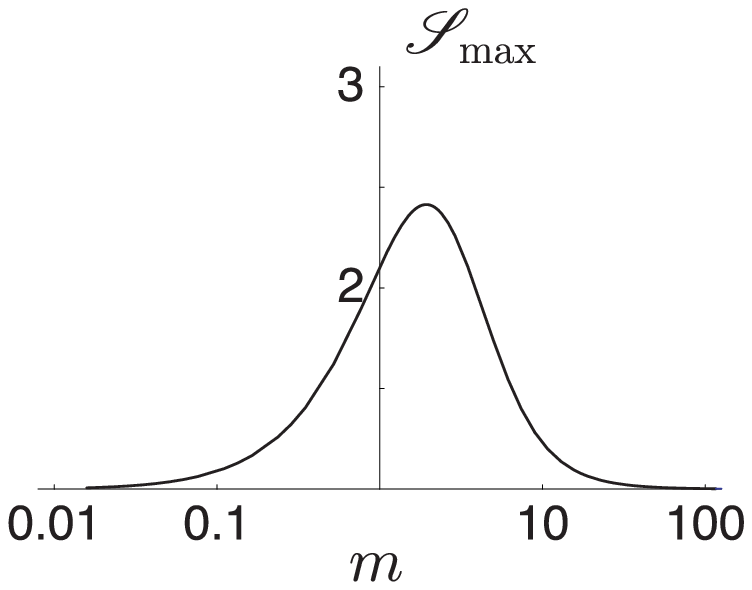} \\
 		(c) & (d) \\
		\end{tabular}
		\caption{Spectral density of two-level lasers with $\gamma=0$, $N=10^5$ and $\alpha=6.32$.  Upper curves are 3D plots of $\spectral(\Omega,J)$. Lower curves give $\spectral_{\mathrm{max}}$ as a function of $m$, the photon number within the cavity. At a fixed $J$, $\spectral_{\mathrm{max}}$ is the maximum value of $\spectral(\Omega,J)$ when $\Omega$ is varied: (a) and (c) $\xi = 1$, Poissonian pump, (b) and (d) $\xi = 0$, quiet pump.}
		 \label{densities}
\end{figure}

The features of $\spectral(\Omega)$ just discussed are illustrated in Fig.~\ref{densities}.  We have chosen two-level laser parameters corresponding to a thresholdless laser ($\gamma=0$), $N=10^5$ atoms, and cavity losses $\alpha=6.32$ identical to those considered for three- and four-level lasers in \cite{chusseau:3levels}.  Figs.~\ref{densities}a and \ref{densities}b verify that two-level lasers are noisy even when spontaneous decay is negligible.  This occurs at a Fourier frequency that depends on the pumping level. The maximum spectral-density value is a function of the number $m$ of photons stored in the cavity. The curves given in Figs.~\ref{densities}c and \ref{densities}d demonstrate that the laser is most noisy when $m$ is on the order of unity. At small
$m$-values there is no phase-coherence between successive wave trains emitted because they do not overlap, and the light is thermal-like. At high $m$-values a regulation mechanism takes place.

\section{Fano factor}
\label{FF}

The intra-cavity photon statistics is characterized by the Fano factor $\fano = \mathrm{var}(m) / ave{m}$.  The variance of $m$ is the integral over frequency of $\spectral_{\Delta m}(\Omega )$, the normalized spectral density of $\Delta m$.  Similar to Sec.~\ref{PDS}, $\spectral_{\Delta m}(\Omega )$ is obtained by solving Eqs.~\eqref{fluctuations}--\eqref{steadyf} for $\Delta m$ instead of $\Delta Q$
\begin{equation}
	\spectral_{\Delta m}(\Omega ) = \frac{1}{m}\frac{\mathscr{V}(\Omega)}{\mathscr{D}(\Omega)} ,
    \label{eq:SDm}
\end{equation}
where
\begin{equation}
	\mathscr{V}(\Omega) =  \left( n_{2} (m+1) + \alpha m \right) \Omega^2 + n_{2}(m + 1)  \gamma \widehat{m} + \alpha m \widehat{m}^2 + J (m + 1)^2 \xi , 
	\label{V}
\end{equation}
and with $\mathscr{D}(\Omega)$ still given by Eq.~\eqref{D}.

The Fano factor is thus
\begin{equation}
	\fano = \int_{-\infty}^{\infty} \spectral_{\Delta m}(\Omega )
	\frac{\mathrm{d}\Omega}{2 \pi} ,
	\label{eq:Fano}
\end{equation}
for which a lengthy analytical expression has been obtained using symbolic computations.  Well above threshold and with $\gamma = 0$, this expression yields \( \fano = \left( 1 + \xi \right) / 2 \), in 
agreement with \cite{arnaud:OQE95}.

\begin{figure}
    \begin{tabular}{cc}
		(a) & (b) \\
		\includegraphics[width=0.45\linewidth]{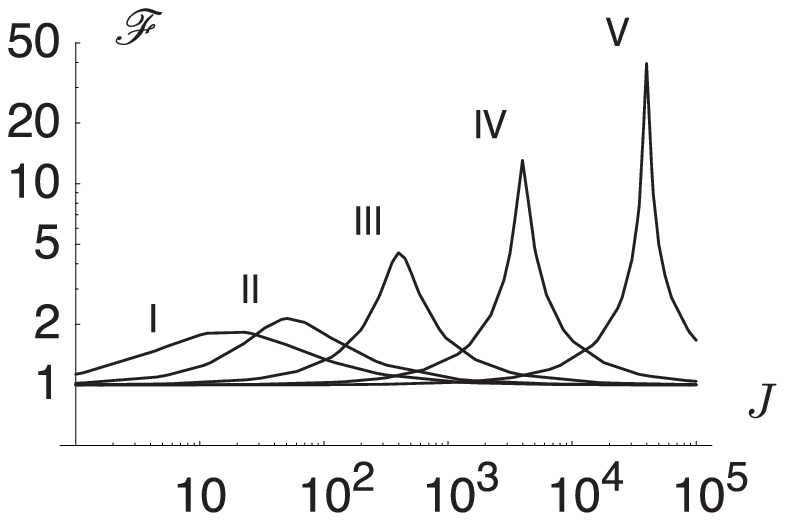} & 
		\includegraphics[width=0.45\linewidth]{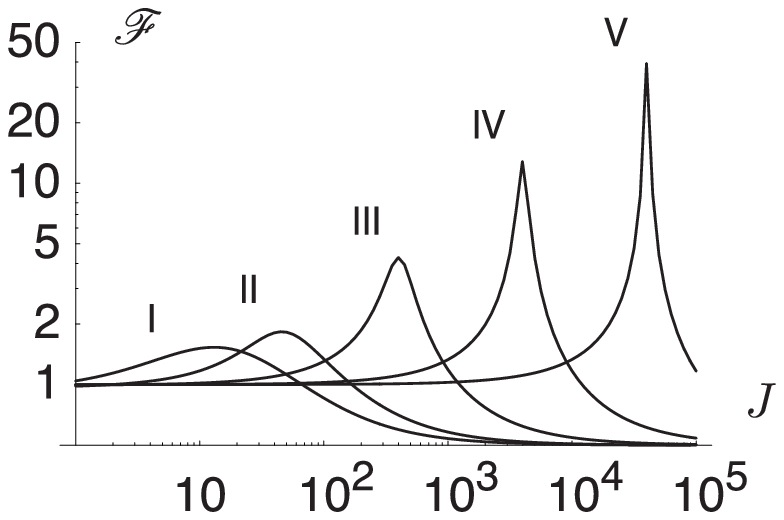} \\
    \end{tabular}
\caption{Fano factor of two-level lasers with parameters $N=10^5$ and $\alpha=6.32$ as a function of the pump rate $J$ and the spontaneous emission rate $\gamma$.  (a) $\xi = 1$, poissonian pump, (b) $\xi = 0$, quiet pump.  (I) $\gamma=0$, (II) $\gamma=6.32$, (III) $\gamma=63.2$, (IV) $\gamma=632$, (V) $\gamma=6325$.} 
\label{fano}
\end{figure}

The Fano factor of two-level lasers is represented as a function of the pumping rate and the spontaneous decay rate in Fig.~\ref{fano}. When $\gamma=0$ a bump is observed on $\fano$ at $J \approx 10$, related to the bump observed on $\spectral$ in Fig.~\ref{densities}: light
statistics is nearly that of thermal light.  The bump sharpens as $\gamma$ increases with a maximum occurring at threshold.  Below threshold the light statistics is thermal-like, so that $\fano \approx \ave{m}+1$. Above threshold Poissonian light statistics ($\fano \approx 1$) is being approached, for the case of a Poissonian pump.

\section{Conclusion}
\label{Conc}

Two-level lasers have been considered within the framework of generalized rate-equations.  Fully analytical expressions of the photo-current spectral density and intracavity field Fano factor were obtained.  It was shown that thresholdless two-level lasers become noisy when the average number of photons in the cavity is on the order of unity. Sub-Poissonian light statistics is accurately described. The present paper should be considered as being partly tutorial, in that it provides more accessible formulas than does a previous paper dealing with four-level atom lasers  \cite{chusseau:3levels}.

\bibliography{Two-level}

\begin{thebibliography}{16}
\expandafter\ifx\csname natexlab\endcsname\relax\def\natexlab#1{#1}\fi
\expandafter\ifx\csname bibnamefont\endcsname\relax
  \def\bibnamefont#1{#1}\fi
\expandafter\ifx\csname bibfnamefont\endcsname\relax
  \def\bibfnamefont#1{#1}\fi
\expandafter\ifx\csname citenamefont\endcsname\relax
  \def\citenamefont#1{#1}\fi
\expandafter\ifx\csname url\endcsname\relax
  \def\url#1{\texttt{#1}}\fi
\expandafter\ifx\csname urlprefix\endcsname\relax\def\urlprefix{URL }\fi
\providecommand{\bibinfo}[2]{#2}
\providecommand{\eprint}[2][]{\url{#2}}

\bibitem[{\citenamefont{Yamamoto et~al.}(1986)\citenamefont{Yamamoto, Machida,
  and Nilsson}}]{yamamoto:PRA86}
\bibinfo{author}{\bibfnamefont{Y.}~\bibnamefont{Yamamoto}},
  \bibinfo{author}{\bibfnamefont{S.}~\bibnamefont{Machida}}, \bibnamefont{and}
  \bibinfo{author}{\bibfnamefont{O.}~\bibnamefont{Nilsson}},
  \bibinfo{journal}{Phys. Rev. A} \textbf{\bibinfo{volume}{34}},
  \bibinfo{pages}{4025} (\bibinfo{year}{1986}).

\bibitem[{\citenamefont{Kennedy and Walls}(1989)}]{kennedy:PRA89}
\bibinfo{author}{\bibfnamefont{T.~A.~B.} \bibnamefont{Kennedy}}
  \bibnamefont{and} \bibinfo{author}{\bibfnamefont{D.~F.} \bibnamefont{Walls}},
  \bibinfo{journal}{Phys. Rev. A} \textbf{\bibinfo{volume}{40}},
  \bibinfo{pages}{6366} (\bibinfo{year}{1989}).

\bibitem[{\citenamefont{Khazanov et~al.}(1990)\citenamefont{Khazanov, Koganov,
  and Gordov}}]{khazanov:PRA90}
\bibinfo{author}{\bibfnamefont{A.~M.} \bibnamefont{Khazanov}},
  \bibinfo{author}{\bibfnamefont{G.~A.} \bibnamefont{Koganov}},
  \bibnamefont{and} \bibinfo{author}{\bibfnamefont{E.~P.}
  \bibnamefont{Gordov}}, \bibinfo{journal}{Phys. Rev. A}
  \textbf{\bibinfo{volume}{42}}, \bibinfo{pages}{3065} (\bibinfo{year}{1990}).

\bibitem[{\citenamefont{Ritsch et~al.}(1991)\citenamefont{Ritsch, Zoller,
  Gardiner, and Walls}}]{ritsch:PRA91}
\bibinfo{author}{\bibfnamefont{H.}~\bibnamefont{Ritsch}},
  \bibinfo{author}{\bibfnamefont{P.}~\bibnamefont{Zoller}},
  \bibinfo{author}{\bibfnamefont{C.~W.} \bibnamefont{Gardiner}},
  \bibnamefont{and} \bibinfo{author}{\bibfnamefont{D.~F.} \bibnamefont{Walls}},
  \bibinfo{journal}{Phys. Rev. A} \textbf{\bibinfo{volume}{44}},
  \bibinfo{pages}{3361} (\bibinfo{year}{1991}).

\bibitem[{\citenamefont{Ralph and Savage}(1991)}]{ralph:PRA91}
\bibinfo{author}{\bibfnamefont{T.~C.} \bibnamefont{Ralph}} \bibnamefont{and}
  \bibinfo{author}{\bibfnamefont{C.~M.} \bibnamefont{Savage}},
  \bibinfo{journal}{Phys. Rev. A} \textbf{\bibinfo{volume}{44}},
  \bibinfo{pages}{7809} (\bibinfo{year}{1991}).

\bibitem[{\citenamefont{Kolobov et~al.}(1993)\citenamefont{Kolobov, Davidovich,
  Giacobino, and Fabre}}]{kolobov:PRA93}
\bibinfo{author}{\bibfnamefont{M.~I.} \bibnamefont{Kolobov}},
  \bibinfo{author}{\bibfnamefont{L.}~\bibnamefont{Davidovich}},
  \bibinfo{author}{\bibfnamefont{E.}~\bibnamefont{Giacobino}},
  \bibnamefont{and} \bibinfo{author}{\bibfnamefont{C.}~\bibnamefont{Fabre}},
  \bibinfo{journal}{Phys. Rev. A} \textbf{\bibinfo{volume}{47}},
  \bibinfo{pages}{1431} (\bibinfo{year}{1993}).

\bibitem[{\citenamefont{Levien et~al.}(1993)\citenamefont{Levien, Collett, and
  Walls}}]{levien:PRA93}
\bibinfo{author}{\bibfnamefont{R.~B.} \bibnamefont{Levien}},
  \bibinfo{author}{\bibfnamefont{M.~J.} \bibnamefont{Collett}},
  \bibnamefont{and} \bibinfo{author}{\bibfnamefont{D.~F.} \bibnamefont{Walls}},
  \bibinfo{journal}{Phys. Rev. A} \textbf{\bibinfo{volume}{47}},
  \bibinfo{pages}{5030} (\bibinfo{year}{1993}).

\bibitem[{\citenamefont{Koganov and Shuker}(2000)}]{koganov:PRA00}
\bibinfo{author}{\bibfnamefont{G.~A.} \bibnamefont{Koganov}} \bibnamefont{and}
  \bibinfo{author}{\bibfnamefont{R.}~\bibnamefont{Shuker}},
  \bibinfo{journal}{Phys. Rev. A} \textbf{\bibinfo{volume}{63}},
  \bibinfo{pages}{015802} (\bibinfo{year}{2000}).

\bibitem[{\citenamefont{Arnaud}(1995)}]{arnaud:OQE95}
\bibinfo{author}{\bibfnamefont{J.}~\bibnamefont{Arnaud}},
  \bibinfo{journal}{Opt. Quantum Electron.} \textbf{\bibinfo{volume}{27}},
  \bibinfo{pages}{63} (\bibinfo{year}{1995}).

\bibitem[{\citenamefont{Arnaud}(2002)}]{arnaud:OQE01}
\bibinfo{author}{\bibfnamefont{J.}~\bibnamefont{Arnaud}},
  \bibinfo{journal}{Opt. Quantum Electron.} \textbf{\bibinfo{volume}{34}},
  \bibinfo{pages}{393} (\bibinfo{year}{2002}).

\bibitem[{\citenamefont{Chusseau et~al.}(2002)\citenamefont{Chusseau, Arnaud,
  and Philippe}}]{chusseau:3levels}
\bibinfo{author}{\bibfnamefont{L.}~\bibnamefont{Chusseau}},
  \bibinfo{author}{\bibfnamefont{J.}~\bibnamefont{Arnaud}}, \bibnamefont{and}
  \bibinfo{author}{\bibfnamefont{F.}~\bibnamefont{Philippe}}
  (\bibinfo{year}{2002}), \eprint{quant-ph/0203029}.

\bibitem[{\citenamefont{Loudon}(1983)}]{loudon}
\bibinfo{author}{\bibfnamefont{R.}~\bibnamefont{Loudon}},
  \emph{\bibinfo{title}{{T}he {Q}uantum {T}heory of {L}ight}}
  (\bibinfo{publisher}{Oxford University Press}, \bibinfo{address}{Oxford},
  \bibinfo{year}{1983}).

\bibitem[{\citenamefont{Chusseau and Arnaud}(2002)}]{chusseau:OQE01}
\bibinfo{author}{\bibfnamefont{L.}~\bibnamefont{Chusseau}} \bibnamefont{and}
  \bibinfo{author}{\bibfnamefont{J.}~\bibnamefont{Arnaud}},
  \bibinfo{journal}{Opt. Quantum Electron.}  (\bibinfo{year}{2002}),
  \bibinfo{note}{to appear}, \eprint{quant-ph/0105078}.

\bibitem[{\citenamefont{Scully and Lamb}(1967)}]{scully:PR67}
\bibinfo{author}{\bibfnamefont{M.~O.} \bibnamefont{Scully}} \bibnamefont{and}
  \bibinfo{author}{\bibfnamefont{W.~E.} \bibnamefont{Lamb},
  \bibfnamefont{Jr.}}, \bibinfo{journal}{Phys. Rev.}
  \textbf{\bibinfo{volume}{159}}, \bibinfo{pages}{208} (\bibinfo{year}{1967}).

\bibitem[{\citenamefont{Golubev and Sokolov}(1984)}]{golubev:JETP84}
\bibinfo{author}{\bibfnamefont{Y.~M.} \bibnamefont{Golubev}} \bibnamefont{and}
  \bibinfo{author}{\bibfnamefont{I.~V.} \bibnamefont{Sokolov}},
  \bibinfo{journal}{Sov. Phys.-JETP} \textbf{\bibinfo{volume}{60}},
  \bibinfo{pages}{234} (\bibinfo{year}{1984}).

\bibitem[{\citenamefont{Machida et~al.}(1987)\citenamefont{Machida, Yamamoto,
  and Itaya}}]{machida:PRL87}
\bibinfo{author}{\bibfnamefont{S.}~\bibnamefont{Machida}},
  \bibinfo{author}{\bibfnamefont{Y.}~\bibnamefont{Yamamoto}}, \bibnamefont{and}
  \bibinfo{author}{\bibfnamefont{Y.}~\bibnamefont{Itaya}},
  \bibinfo{journal}{Phys. Rev. Lett.} \textbf{\bibinfo{volume}{58}},
  \bibinfo{pages}{1000} (\bibinfo{year}{1987}).

\end{thebibliography}

\end{document}